\documentclass[preprint]{elsarticle}
\usepackage{graphicx,url,hyperref,microtype,subcaption,graphbox,amsmath}
\usepackage{latexsym,courier,upgreek,xcolor,rotating,ulem}
\usepackage[onehalfspacing]{setspace}
\hypersetup{colorlinks,citecolor=blue,linkcolor=red,urlcolor=green}
\usepackage[displaymath,mathlines]{lineno}
\usepackage[margin=2cm]{geometry}
\modulolinenumbers[5]
\biboptions{numbers,sort&compress}

\begin{document}

\begin{frontmatter}
  
  \title{Correlation between avalanches and emitted energies during fracture with variable stress release range}
  
  \author[add1]{Narendra K. Bodaballa}
  \ead{narendrakumar_b@srmap.edu.in}
  
  \author[add1]{Soumyajyoti Biswas}
  \ead{soumyajyoti.b@srmap.edu.in}
  
  \author[add2]{Subhadeep Roy}
  \ead{subhadeeproy03@gmail.com}
  
  \address[add1]{Department of Physics, SRM University - AP, Andhra Pradesh 522502, India.}
    
  \address[add2]{PoreLab, Department of Physics, Norwegian University of Science and Technology, N-7491 Trondheim, Norway.}

  \begin{abstract}
    We observe the failure process of a fiber bundle model with a variable stress release range, $\gamma$, higher the value of $\gamma$ lower the stress release range. By tuning $\gamma$ from low to high, it is possible to go from the mean-field (MF) limit of the model to local load sharing (LLS) where local stress concentration plays a crucial role. In the MF limit, the avalanche size $s$ and energy $E$ emitted during the avalanche are highly correlated producing the same distribution for both $P(s)$ and $Q(E)$: a scale-free distribution with a universal exponent -5/2. With increasing $\gamma$, the model enters the LLS limit. In this limit, due to the presence of local stress concentration such correlation $C(\gamma)$ between $s$ and $E$ decreases where the nature of the decreases depends highly on the dimension of the bundle. In 1d, the $C(\gamma)$ stars from a high value for low $\gamma$ and decreases towards zero when $\gamma$ is increased. As a result, $Q(E)$ and $P(s)$ are similar at low $\gamma$, an exponential one, and then $Q(E)$ becomes power-law for high-stress release range though $P(s)$ remains exponential. On the other hand, in 2d, the $C(\gamma)$ decreases slightly with $\gamma$ but remains at a high value. Due to such a high correlation, the distribution of both $s$ and $E$ is exponential in the LLS limit independent of how large $\gamma$ is.
  \end{abstract}

  \date{\today}
  
  
\end{frontmatter}


\makeatletter
\def\ps@pprintTitle{%
 \let\@oddhead\@empty
 \let\@evenhead\@empty
 \def\@oddfoot{}%
 \let\@evenfoot\@oddfoot}
\makeatother


\section{Introduction}
Disordered materials, when subjected to external stress, goes through local breakdowns that eventually emerge as a catastrophic fracture when the load exceeds a critical value. These breaking events, called avalanches, are usually detected as accoustic emissions \cite{phys_rep}. Systematic statistical analysis of the time series of these events have led to considerable insight into the failure dynamics of  disordered samples across scales. Particularly, the scale free nature of the size distribution of these avalanches have motivated a `critical phenomena' description of fracture processes. From laboratory scale experiments to 
earthquake statistics, the Guttenberg-Richter-like law is widely and accurately verified \cite{rmp_2012,lucilla}. 

The advantage of having a critical phenomena description is that the dynamics is expected to be independent of the microscopic details of the individual system studied, and will depend of a few parameters such as the dimensionality of the system, the interaction range, order parameter definition etc. (see e.g., \cite{bonamy}). This means, simplified models with a very few parameters should be able to reproduce the `critical description' of breadown phenomena. Indeed, there have been many such attempts. Some of the prominent of such models include the fiber bundle model \cite{Pierce}, the random fuse model \cite{fuse_model}, Burridge Knopoff model for earthquakes \cite{bk_model}, and so on. There are also multiple efforts in molecular dynamics simulations to obtain the scale free avalanche statistics.

While in experiments avalanches are usually detected as acoustic emissions, the analogue in simulations is not unique. In the fiber bundle model, which is an ensemble of discrete elements having different individual failure threshold, an avalanche is often defined as the number of elements breaking due to a small increase in the global load on the system. When the individual elements -- fibers -- are assumed to exhibit some stress-strain response, the `energy' and `size' of an avalanche are, in general, distinct quantities. While the energy size distributions for the fiber bundle are also studied, it is very often used interchangably with avalanche size. In the mean field or the equal load sharing limit of the model, these two quantities differ only by a factor, hence their distribution functions are the same. But a departure from the mean-field limit would, in general, not keep this equivalence. In particular, it was noted \cite{front_20} that in the extreme limit i.e., the nearest neighbor load sharing, the
avalanche size distribution is exponential but the energy distribution is a power law. 

In fiber bundle model, the mode of failure depends mainly on two factors: the strength of disorder and the range of stress release \cite{rr15,r17,brr15,rsr17}. Such interplay between disorder and stress release range not one affects the failure mode but also influences the spatial correlation during the failure process \cite{brr15,rsr17,srh20,r21a,r21b}. the The range of stress release has the two extreme limits as the equal and nearest neighborhood sharing. However, real situations are in between these two limits. Therefore, it is crucial to know to what extent the equivalence between avalanche and energy sizes is valid in the model. In other words, if comparisons are to be made with experiments, which usually measures emitted energies and not the size of avalanches, then for any localization of the load redistribution, it needs to be investigated whether avalanche size or energy is the appropriate quantity to look at or upto what extent these two are equivalent. In this work, we study the fiber bundle model in one and two dimensions with variable range of load sharing. We look at the distributions of the avalanche size and energy. We also look at the correlation measures for the time series of avalanche sizes and energy as a function of the range of stress release. 


\section{Description of Fiber Bundle Model}

Fiber bundle model has been proven to be very useful yet very simple model to study failure process in disordered systems. It has gained a lot of attention among engineers, material scientists and physicists after its introduction by Pierce in 1926 \cite{Pierce}. A conventional fiber bundle model of size $L$ consists of $L$ parallel Hookean fibers places vertically between two supporting clamps. The clams are pulled apart by a force $F$, creating a stress $f=F/L$ per fiber. Disorder is introduced within the model as fluctuation of strength values of individual fiber. In the present work, such strength values ($h$) are chosen from a uniform distribution with mean at 0.5 and width $2\delta$. 
\begin{equation}\label{eq1}
\rho(h) \sim \begin{cases}
    \displaystyle\frac{1}{2\delta},  & (0.5-\delta \le h \le 0.5+\delta) \\
    0  & ({\rm otherwise})
  \end{cases}
\end{equation}
$\rho(h)$ is the probability of getting a threshold $h$. The half-width $\delta$ is the measure of disorder strength here, higher the $\delta$ value higher the strength of disorder. A certain fiber $i$ will break irreversibly if the applied stress $\sigma(i)$ on that fiber crosses its threshold value $h(i)$. The stress of broken fibers is then redistributed either globally among all surviving fibers (global load sharing or GLS scheme) \cite{Pierce, Daniels} or among the surviving nearest neighbors only (local load sharing or LLS scheme) \cite{Phoenix,Smith,Newman,Harlow2,Harlow3,Smith2}. This is the two extreme limit of the model. The GLS limit is the case when the clamps, supporting the fibers, are extremely rigid and the response of the broken fiber can travel throughout the bundle. On the other hand, we enter the LLS limit when if the clamps are very soft and the effect of a certain rupture is experienced in its neighborhood only.     

Here we have adopted a generalized version of fiber bundle model where the stress release range during the failure process can be tuned. The algorithm for such redistribution is discussed next. If $\sigma_i$ is the stress of the broken fiber $i$, then for a certain fiber $j$, the local stress profile after redistribution will obey the following rule: 
\begin{equation}\label{eq2}
\sigma_j \rightarrow \sigma_j + \displaystyle\frac{r_{ij}^{-\gamma}}{Z} \sigma_i
\end{equation}
$r_{ij}$ is the distance between the fibers $i$ and $j$. $Z$ is the normalization factor given by
\begin{equation}\label{eq3}
Z = \displaystyle\sum_{i,k} r_{ik}^{-\gamma}
\end{equation}
where $k$ runs over all intact fibers. Such $\gamma$-dependent stress redistribution has been explored earlier \cite{hmkh02,brr15} in the context of fiber bundle model. With the present rule, a high $\gamma$ stands for a low stress release range and hence closer to LLS scheme. On the other hand, for low $\gamma$, the model enters the GLS or mean-field limit. Earlier studies pointed out a critical $\gamma_c$ for both 1d \cite{brr15} and 2d \cite{hmkh02} fiber bundle model around which this transition from LLS to GLS limit takes place. After such redistribution, other fibers can break due to the elevated local stress profile and the stress is again redistributed. This starts an avalanche at the same external stress. The size $s$ of an avalanche is the number of fibers broken during the process. The energy emitted during an avalanche of size $s$ is represented as follows: 
\begin{equation}\label{eq4}
E(s) = \displaystyle\sum_{1}^s \displaystyle\frac{1}{2}h_i^2
\end{equation}
where $h_i$ are the threshold values of the $s$ broken fibers during the avalanche. An average energy $\langle E \rangle$ can be calculated then by averaging over all $E$ values corresponding to a certain avalanche size $s$. The total bundle might break through a single avalanche, or it can come to a stable state after an avalanche. In the later case, we increase the applied stress just to break the next weakest fiber and start a new avalanche. This process goes on until all fibers are broken suggesting the global failure of the bundle.          


\section{Numerical Results}

Numerically we have studied both 1d and 2d fiber bundle model with a variable stress release range $\gamma$. The system sizes vary from $10^3$ to $10^5$ for 1d and from $50 \times 50$ to $150 \times 150$ for 2d. The strength of disorder is kept at an intermediate value ($\delta=0.5$) so that the avalanches are observed properly. Earlier studies showed that a very low or high $\delta$ values will cause either brittle like abrupt failure or failure by stress increment only \cite{rr15,r17,brr15,rsr17}. In the former case, we obtain a single avalanche of size $L$, while for the later $L$ avalanches will be observed, each of size $1$. The value of $\gamma$ has been increases continuously from 0 to 8, which allows us to explore the whole region, ELS to LLS, during the failure process.

We start by observing the relation between $s$ and $\langle E \rangle$ with a continuous variation in $\gamma$. This will give us an idea about the correlation that the avalanche size and the energy has in between. Such correlation is then explored explicitly through a correlation function $C(\gamma)$ and compared with $s$ vs $\langle E \rangle$ behavior. Finally the distribution of avalanches and emitted energies are shown and connection between them is established through the correlation function. \\


\subsection{Relation between $s$ and $E$}

\begin{figure*}[ht]
\centering
\includegraphics[width=15cm, keepaspectratio]{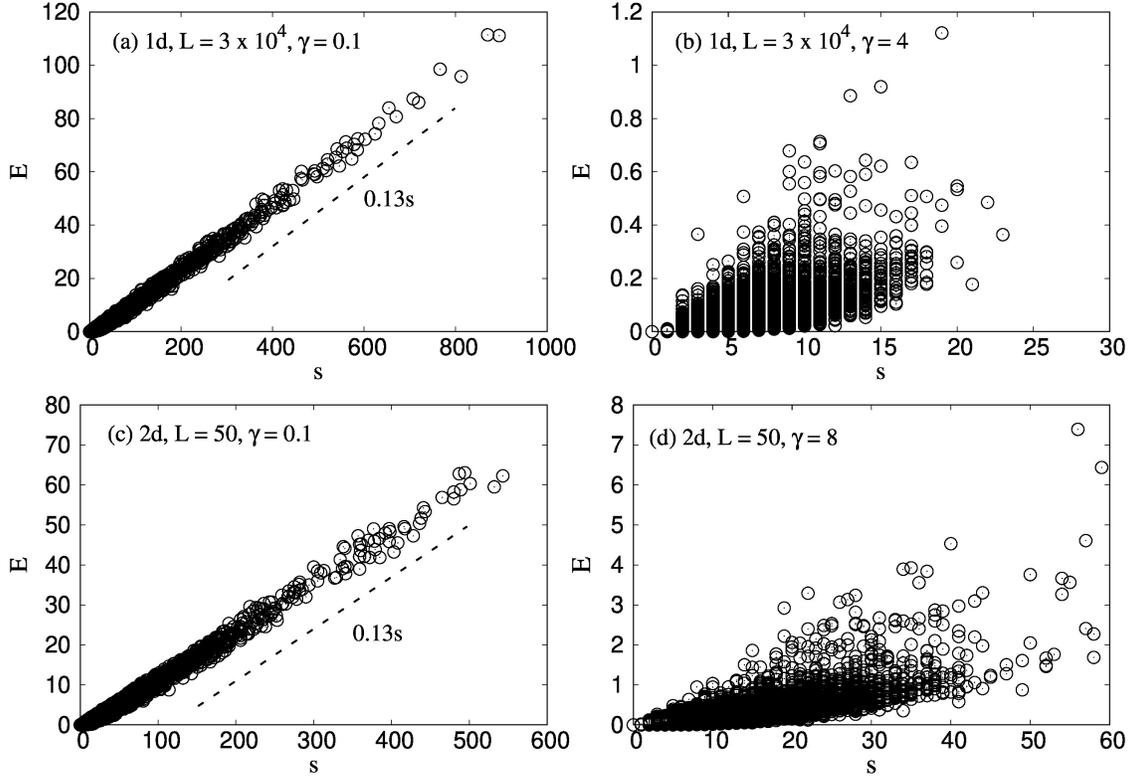}  
\caption{The figure shows the variation of avalanche size $s$ with corresponding emitted energy $E$ for (a) $\gamma=0.1$ and (b) $4.0$ in case of a 1d bundle and for (c) $\gamma=0.1$ and (d) $8.0$ in case of a 2d bundle. There is high correlation between $s$ and $E$ for low $\gamma$ (ELS limit). The plot is rather scatted with a relatively low correlation when $\gamma$ is high (LLS limit). Though this trend is same for both dimensions, the scatter seems less as we go to the higher dimension.}
\label{s_vs_E}
\end{figure*}

Figure \ref{s_vs_E} shows the variation of avalanche size $s$ and energy emitted $E$ during that avalanche. The emitted energy for a certain avalanche is defined by equation \ref{eq4}. We have explored this $s$ vs $E$ behavior for $\gamma=0.1$ and $8.0$. For the former value of $\gamma$, the model is in the mean-field limit while for the later one due to a high value the model is close to the LLS scheme. The upper panel of figure \ref{s_vs_E} shows the results for one dimension and the lower one for 2d FBM. The disorder strength $\delta$ in both dimensions is kept fixed at 0.5. 

We observe that when $\gamma$ is low, $E$ increases with $s$ in a linear manner. A recent article \cite{front_20} has already discussed such correlation between avalanche size and emitted energy in the mean-field limit of the model, which we obtain here by keeping a high range (hence low $\gamma$) of stress relaxation. On the other hand, when $\gamma$ is high, the plot of $s$ vs $E$ shows a scatter with relatively lower correlation between then. Such decrease in correlation with $\gamma$ is observed for both one and two dimensions of the model. We have a more quantitative discussion of the correlation observed here at the end of the article.   

\begin{figure}[ht]
\centering
\includegraphics[width=8.0cm, keepaspectratio]{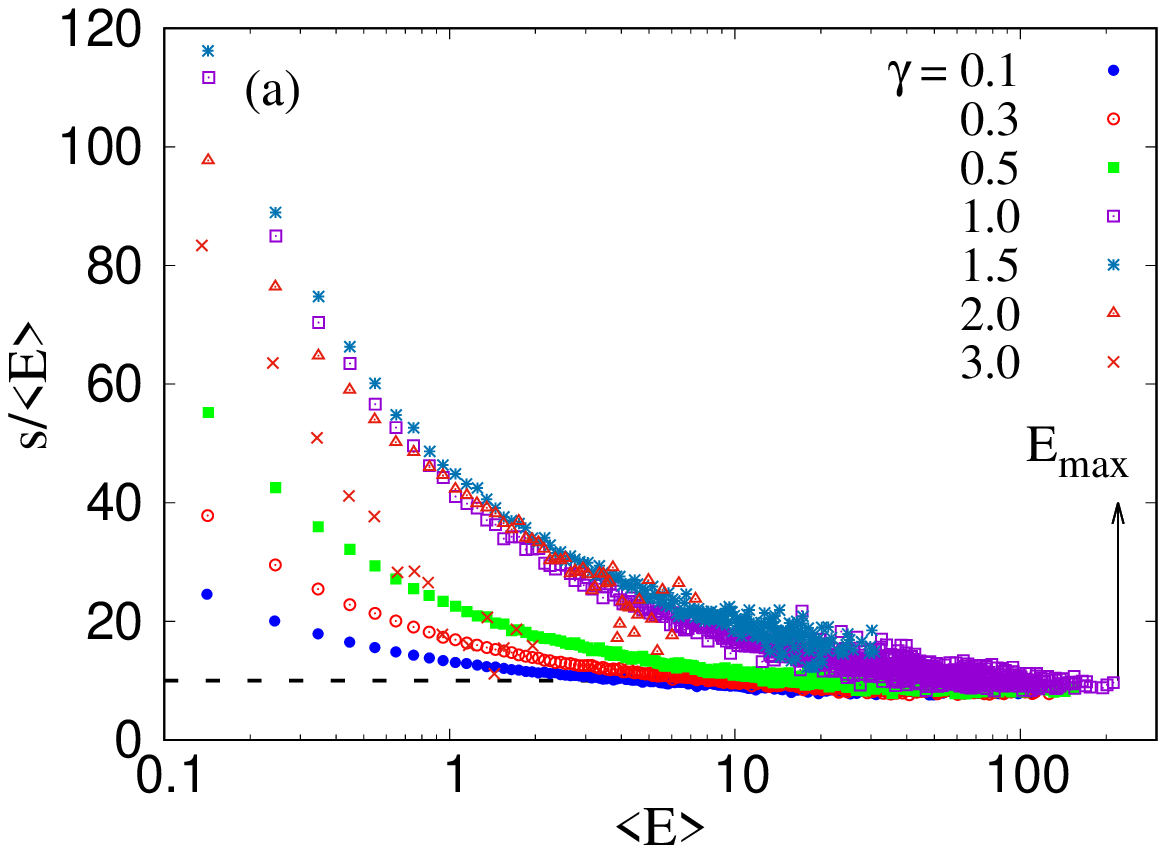} \includegraphics[width=8.0cm, keepaspectratio]{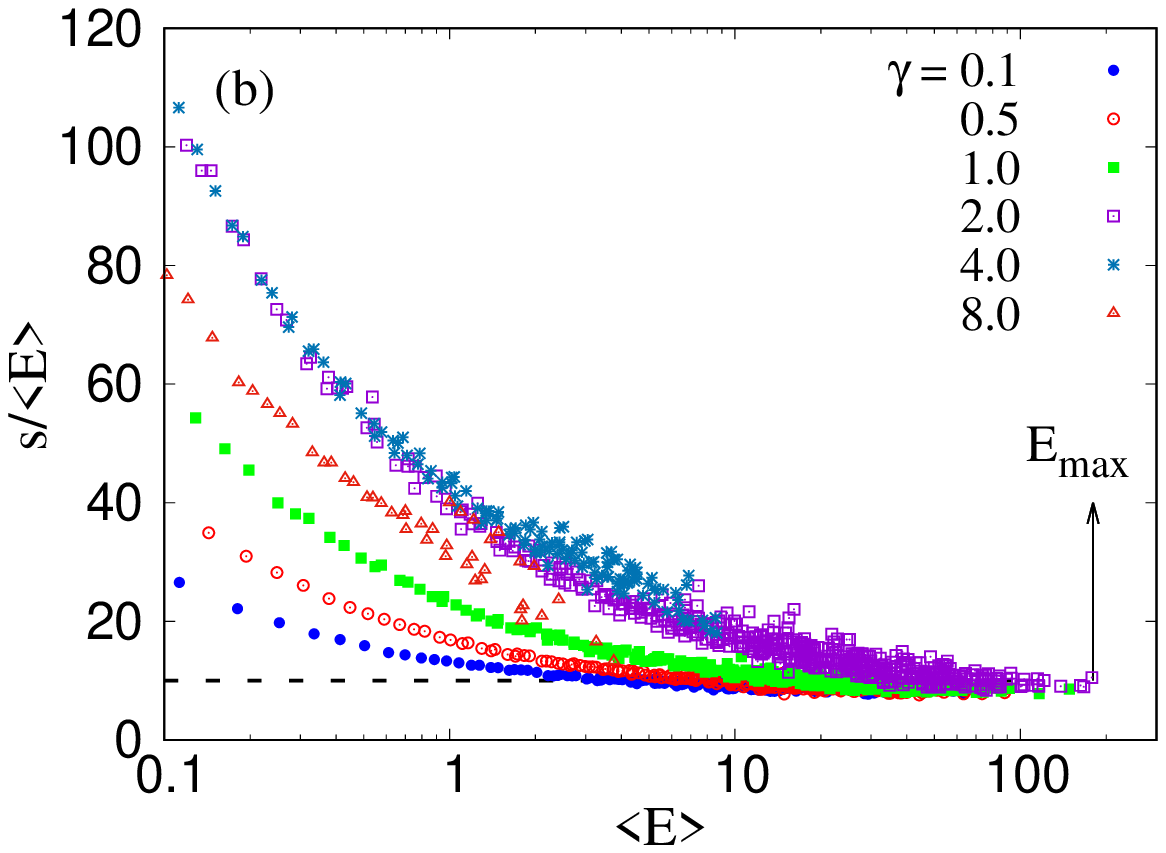}  
\caption{Variation of $s/\langle E \rangle$ with $\langle E \rangle$ for (a) 1d and (b) 2d FBM with $\gamma$ values varying in between 0.1 and 3.0 for the former case and between 0.1 and 3.0 for the latter. The horizontal dotted line, in both figures, are the locus of $\langle E \rangle \sim s$. For low $\gamma$ (ELS limit), $s/\langle E \rangle$ meets the dotted line for high $\langle E \rangle$. For high $\gamma$, the model breaks before $s/\langle E \rangle$ meets the dotted line. The maximum value of $\langle E \rangle$ during the failure of the bundle is denoted as $E_{max}$.}
\label{s_vs_Eave}
\end{figure}

Figure \ref{s_vs_Eave} shows the variation of avalanche size $s$ with average energy $\langle E \rangle$ corresponding to $s$. $\langle E \rangle$ is calculated by averaging over all energy values that is associated with a certain avalanche size $s$. Figure \ref{s_vs_Eave} shows that $s/\langle E \rangle$ starts from a high value for low $\langle E \rangle$ and gradually decreases as $\langle E \rangle$ increases. The horizontal dotted line has a constant value independent of $\langle E \rangle$ suggesting the average energy $\langle E \rangle$ grows linearly with $s$. When $\gamma$ value is low, we observe $\langle E \rangle \sim s$ when the average emitted energy is high with a non-linear initial part corresponding to low $\langle E \rangle$. This happens due to the fact that at low $\gamma$, the model is in the mean-field limit and a high correlation exists between $s$ and $E$ here. A visual representation of this high correlation at low $\gamma$ is already provided in figure \ref{s_vs_E}. 
 
\begin{figure}[ht]
\centering
\includegraphics[width=8.0cm, keepaspectratio]{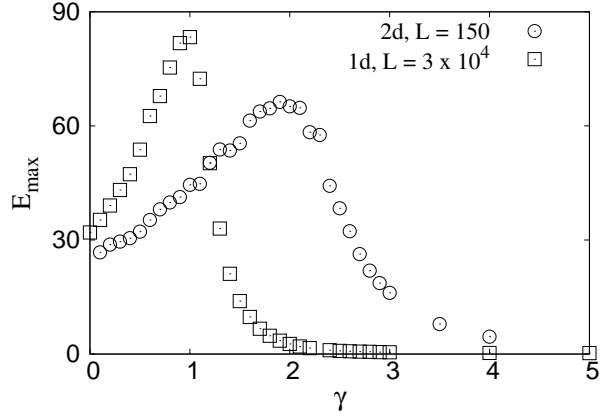}
\caption{Variation of $E_{max}$, maximum possible energy emitted during the failure of the bundle, with the stress release range $\gamma$. The result for 1d and 2d FBM are shown with system size $3 \times 10^4$ and $150$ respectively. For both dimensions, $E_{max}$ shows a peak around $\gamma_c$ and decreases on both sides.}
\label{Emax_vs_gamma}
\end{figure}

On the other hand, for high $\gamma$, the bundle breaks long before we observe linear behavior between $s$ and $\langle E \rangle$. Only the initial non-linear part is observed in this case. The fact that  $\langle E \rangle$ is not linear with $s$ for high $\gamma$ is also reflected by the scattered behavior of $E$ with $s$. We will again come back to this linear and non-linear relationship between $s$ and $\langle E \rangle$ while discussing the distributions of avalanches and emitted energies.

Figure \ref{Emax_vs_gamma} shows how the maximum energy $E_{max}$ emitted during the failure process. We have excluded the energy corresponding to the final avalanche while calculating $E_{max}$. For both 1d and 2d, $E_{max}$ shows a non-monotonic behavior. $E_{max}$ has a low value at both low and high $\gamma$. At low $\gamma$, as the model is closer to the mean field limit, the stress of the broken fiber is redistributed upto to a long distance and a large amount of fibers get a significant part of the redistributed stress. On the other hand, when $\gamma$ is high, most of the redistributed stress is absorbed by the neighboring fibers of the broken one and the amount which is redistributed among others are not significant enough to rupture that fiber. Under such circumstances, most of the bundle breaks abruptly in the final avalanche only. Since we are excluding the final burst, the maximum of energy corresponding to the rest of the avalanches are observed to be smaller as very few avalanches of small sizes takes place prior to global failure. Close to the critical value $\gamma_c$, $E_{max}$ has a maximum value where dynamics of the model is balanced such a way that the final avalanche is not very large and there are many large events producing high energy emission before global failure. The peak shifts to higher values when the system size is increased. This is due to the fact that a larger avalanche prior to global failure will be accessible if the size of the bundle is itself large. Though the $\gamma$ value corresponding to the peak remains unchanged independent of the size of the system.\\


\subsection{Study of the distributions $P(s)$ and $Q(E)$}

\begin{figure}[ht]
\centering
\includegraphics[width=15cm, keepaspectratio]{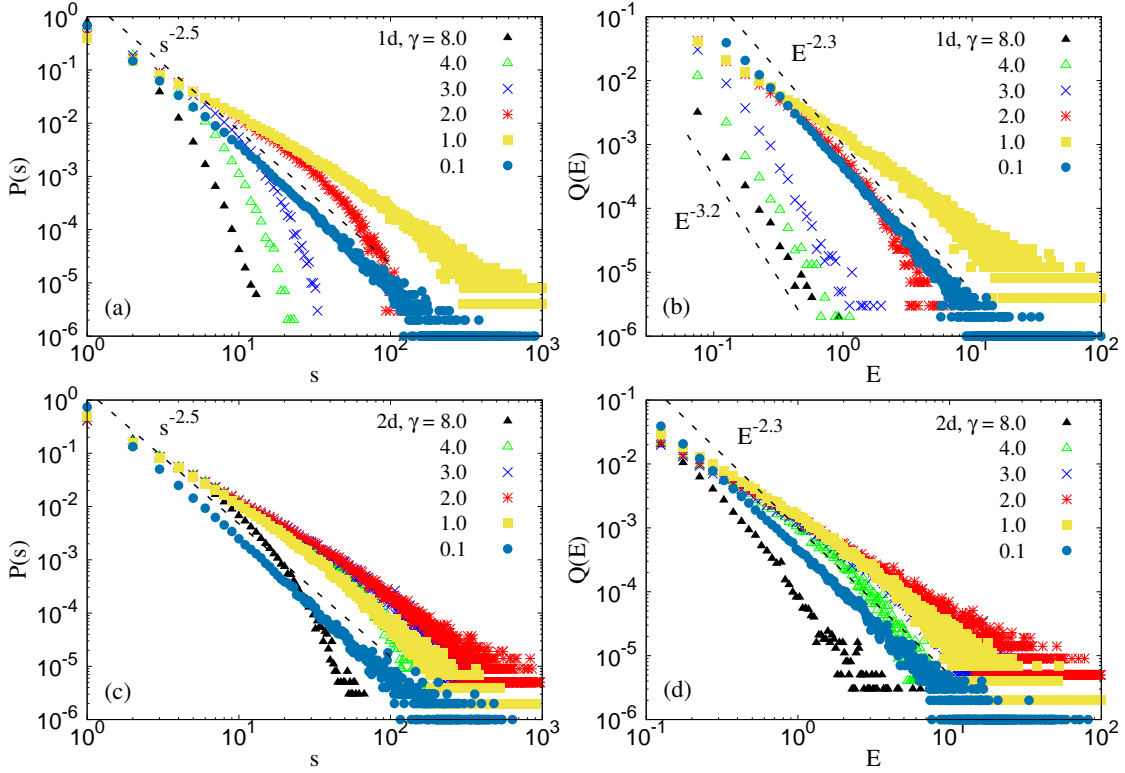} 
\caption{Distribution of the avalanches $P(s)$ and and energies $Q(E)$ with $s$ and $E$ respectively when $\gamma$ is varied in between 0.1 and 8.0. (a) and (b) shows the distributions $P(s)$ and $Q(E)$ in 1d. The same for 2d is shown in (c) and (d). For both 1d and 2d, $P(s)$ is an power law distribution below $\gamma_c$ and an exponential distribution above $\gamma_c$. $Q(E)$, for 1d, is scale-free for $\gamma<\gamma_c$, becomes exponential in the intermediate $\gamma$ value and becomes scale free again when $\gamma$ is high. On the other hand, for 2d FBM, $Q(E)$ is scale free below $\gamma_c$. Beyond $\gamma_c$, it becomes exponential and remains the same even at high $\gamma$.}
\label{s_E_distributions}
\end{figure}

In this section, we will discuss how the distribution of avalanches and energies behave as the stress release range $\gamma$ is varied. Upper panel of figure \ref{s_E_distributions} shows the results for 1d while the results in the lower panel corresponds to 2d FBM. We denote distribution of avalanches by $P(s)$ and distribution of energies by $Q(E)$.  

The distributions are already explored for 1d LLS fiber bundle model in a recent paper \cite{front_20} for two extreme cases, ELS and LLS limit. Here the study is extended to a general $\gamma$ value where previous extreme limits are observed by setting a low and a high $\gamma$ respectively. We observe the following behavior for $P(s)$ and $Q(E)$ with a variation in $\gamma$. 
\begin{itemize}
\item For low $\gamma$ (ELS limit), both $P(s)$ and $Q(E)$ are scale-free distributions with an universal exponent $-2.5$. This is shown in figure \ref{s_E_distributions}(a) and (b) with $\gamma=0.1$. This is due to the fact that in this limit, $\langle E \rangle \sim s$, and the distribution of $E$ should be equal to the distribution $s$ which is a scale-free distribution with exponent $-2.5$ in the ELS limit \cite{hh92}. 
\item On the other hand, when $\gamma$ is very high $Q(E)$ is a scale-free distribution in spite of the fact that the distribution $P(s)$ is an exponential \cite{khh97} here (shown in figure \ref{s_E_distributions}(a) and (b) with $\gamma=8.0$). This is due to the lack of correlation between $s$ and $E$ which is presented in figure \ref{s_vs_E} and \ref{s_vs_Eave}. 
\item For an intermediate $\gamma$ value, both $P(s)$ and $Q(E)$ are exponential distributions. The possible reason could be the fact that the $\gamma$ value where $P(s)$ changes from power law to exponential is not the same $\gamma$ value where the correlation between $s$ and $E$ decreases drastically. This will be clear when we will quantitatively discuss the correlation function. Then there will be an window of $\gamma$ where the correlation is high but $P(s)$ is exponential, producing an exponential distribution of $Q(E)$. 
\end{itemize} 
The behavior in 2d is slightly different than that of 1d FBM. The distributions $P(s)$ and $Q(E)$ are scale-free for low $\gamma$ in case of 2d as well. Both becomes exponential for an intermediate $\gamma$. The only difference is, unlike 1d, $Q(E)$ does not become scale-free again when $\gamma$ is very high. The reason might be as the dimension is increased, the model goes closer to the mean-field limit (as suggested by Sinha et.al \cite{skh15}) and the correlation between $s$ and $E$ will be higher than 1d. Due to a relatively higher correlation, the nature of $P(s)$, which is exponential, is reflected through $Q(E)$ as well. 


\subsection{Correlation Function}

Finally, we reach to a point where we can discuss the correlation between avalanche $s$ and energy emitted $E$ quantitatively. For this, we have adopted the Pearson correlation function and defined a correlation coefficient $C(\gamma)$ given below,
\begin{equation}
C(\gamma) = \displaystyle\frac{n\left(\displaystyle\sum s_n.E_n\right) - \left(\displaystyle\sum s_n\right)\left(\displaystyle\sum E_n\right)}{\sqrt{\left[n\displaystyle\sum s_n^2 - \left(\displaystyle\sum s_n\right)^2\right]\left[n\displaystyle\sum E_n^2 - \left(\displaystyle\sum E_n\right)^2\right]}}
\end{equation}  
where $n$ is total number of avalanches excluding the final one during the failure of the bundle. $s_1$, $s_2$, $\cdots$, $s_n$ are burst sizes of 1st, 2nd, $\cdots$, nth avalanche. Corresponding emitted energies are given as $E_1$, $E_2$, $\cdots$, $E_n$. 

\begin{figure}[ht]
\centering
\includegraphics[width=8.7cm, keepaspectratio]{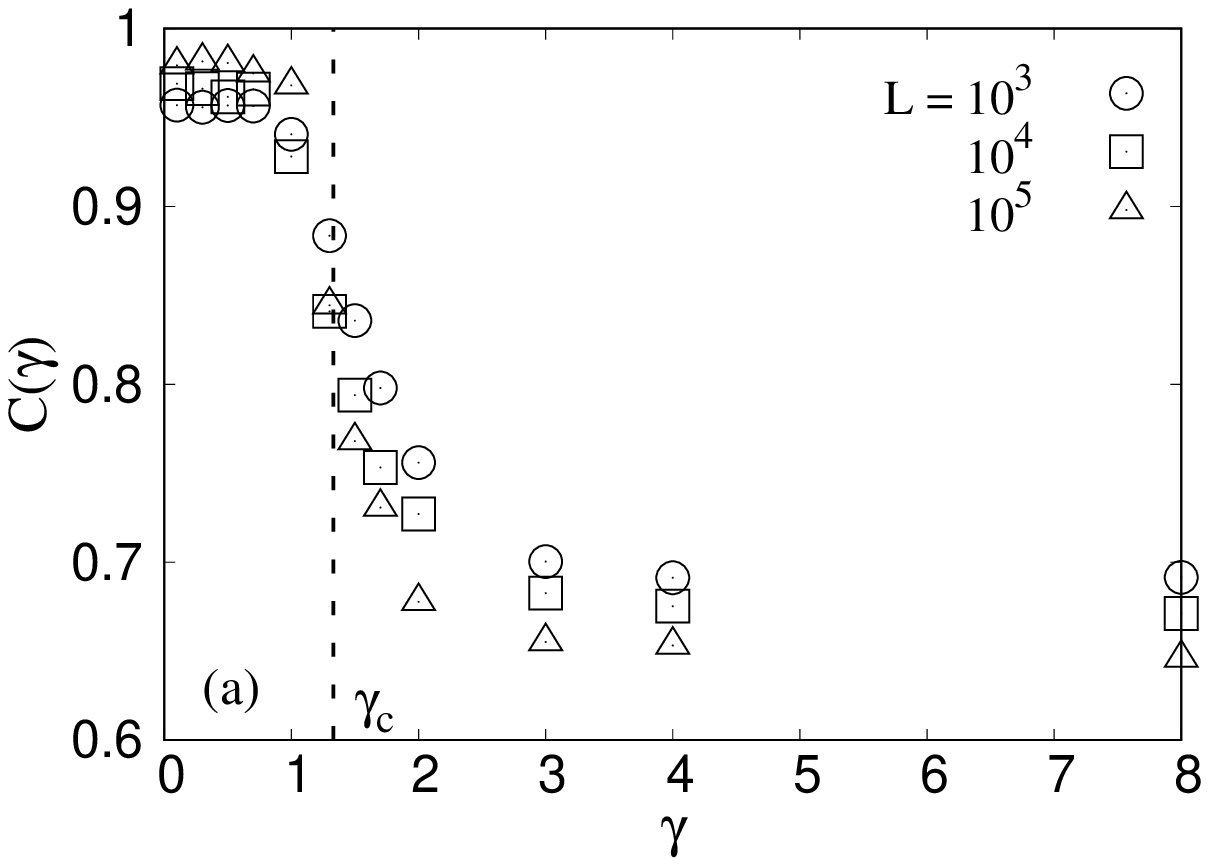} \includegraphics[width=8.7cm, keepaspectratio]{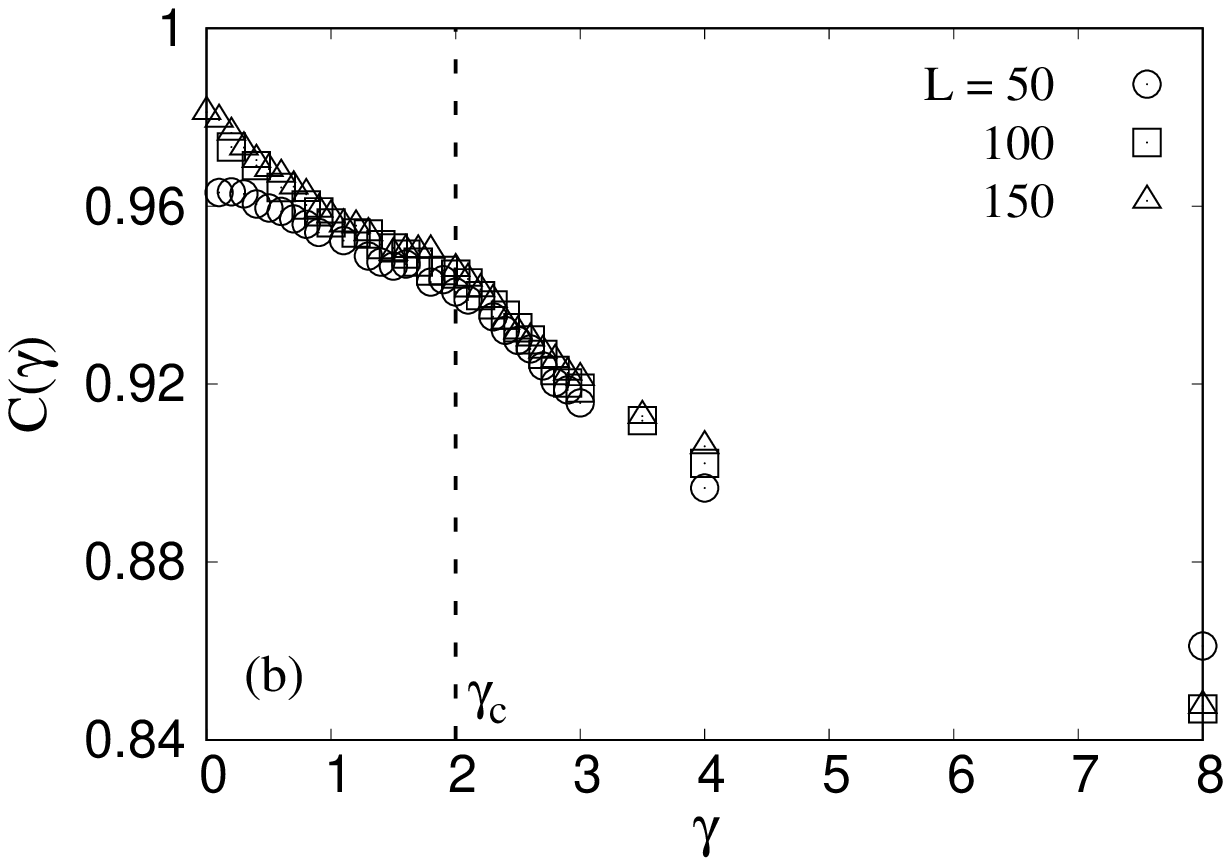} 
\caption{Variation of the correlation function $C(\gamma)$ is shown with $\gamma$ for (a) 1d and (b) 2d fiber bundle model. For $\gamma<\gamma_c$, $s$ and $E$ has high correlation among themselves which was visible in figure \ref{s_vs_E} as well. For $\gamma>\gamma_c$, $C(\gamma)$ falls to a lower value suggesting low $s-E$ correlation. The results for 1d is shown for $L=10^3$, $10^4$ and $10^5$ while the results for 2d are shown for $L=50$, $100$ and $150$. At high $\gamma$, $C(\gamma)$ has a higher value in case of two dimensions.}
\label{s_E_correlation}
\end{figure}

Figure \ref{s_E_correlation} shows how the correlation function $C(\gamma)$ behaves as $\gamma$, the stress release range is increases. The left and the right figure shows the results for 1d and 2d FBM respectively. When $\gamma$ is low $C(\gamma)$ has a value closer to 1.0. This supports our early claim that $s$ and $E$ has a high correlation at low $\gamma$, which corresponds to the ELS or mean-field limit of the model. Similarly, at high $\gamma$, where the model is in the LLS limit, a relatively smaller value of $C(\gamma)$ suggests lesser correlation between $s$ and $E$. This decrease of $C(\gamma)$ takes place around the critical value $\gamma_c$ around which the ELS to LLS transition is observed. When the size of the bundle is increased, $C(\gamma)$ shifts to a higher value at low $\gamma$ and to a lower value at high $\gamma$, suggesting a sharper decay of $C(\gamma)$ around $\gamma_c$. Such effect of system size is more visible in 1d due to higher range of sizes accessible here. Doing the same for 2d will computationally much more costly. Also, notice that the value of $C(\gamma)$ in the LLS limit is higher for 2d as the increase in dimension brings the model closer to the mean-field limit.   


\section{Discussion \& conclusion}
In the fiber bundle model, the failure mode depends on the range of stress release, once the disorder distribution is fixed in a moderate range. The reason for the different modes of failures is the competition of damage progression due to stress localization near an already damaged region and vulnerable spots at some distance from the damaged region. While the first mechanism promotes a nucleation driven failure, the second leads towards a percolative damage. The critical range that separates these two limits have been studied elsewhere \cite{brr15,rsr17}. The underlying mechanism is not only valid for the fiber bundle model \cite{brr15,rsr17,srh20,r21a,r21b}, but is also seen in the random fuse model \cite{mohaha12,sch} and also in molecular dynamics simulations \cite{yuta}. The present study is also a manifestation of these two competing effects. Indeed, for a wide range of stress release, load increases on almost all elements and hence the weaker among those elements break first. Given that there is no spatial correlation between the failure thresholds of the fibers, the damage is percolative. Additionally, the energy released, which is essentially the square of the threshold value, remains small. As the stress release is more and more localized, the maximum energy emitted eventually reaches a peak, which corresponds to the critical value of the localization parameters, beyond which the avalanche size distribution is known to change to exponential form. It is also known that the moments of the cluster size distributions of the damages reach a peak there \cite{hidalgo}. This is the optimal point  where the stress release is sufficiently wide so as to sample the vulnerable fibers and yet sufficiently localized such that the stress concentration on those fibers are high enough. The maximum sizes are therefore seen around this region (see Figs. \ref{s_vs_Eave}, \ref{Emax_vs_gamma}). Upto this point, therefore, the `equivalence' of the avalanche size and energy are valid ($s$ and $E$ reach a linear relation). Beyond this point, these two quantities are to be considered separately. 

Specifically, we have plotted $s$ vs $E$ in Fig. \ref{s_vs_E}, where a clear linearity is observed for wide range of stress redistribution, but for localized ranges, the linearity is not present. The correlation between the two quantities drop from one as the stress release is localized. However, a positive correlation always remains, which is anyway expected. But there could be individual events that are of larger sizes, but still can emit relatively less energy. In the distribution functions the differences are more apparent. For very high values of $\gamma$, the avalanche size distribution always remain exponential, but the energy distribution can show power law behavior with a higher value of exponent (see Fig. \ref{s_E_distributions}). This highlights the fact that in the localized load sharing, one can see power law distributions in energy, which is what is usually measured in the experiments, while the size distributions can already shift to exponentials. 

In conclusion, the interchangability of the avalanche size and energy in he fiber bundle model is only valid when the stress release range in very wide. Specifically, this equivalence is not valid in the nucleation failure mode. However, not only that the correlations between the two quantities are less, the size distribution of the energy can still show a scale free behavior where the avalanche size distribution may not. Gievn that in real materials stress release range is always somewhat local, the observations here can explain the simultaneous existence of damage localization and scale free distributions of emitted energies.


\section{Acknowledgment}

SR supported by the Research Council of Norway through its Centres of Excellence funding scheme, project number 262644.



\end{document}